\documentclass[prd,twocolumn,groupedaddress,showpacs,nofootinbib]{revtex4}
\usepackage{graphicx}
\usepackage{amsmath}
\usepackage{epsfig}

\begin{document}

\title{Radiative Neutrino Mass, Dark Matter and Leptogenesis}

\author{Pei-Hong Gu$^{1}_{}$}
\email{pgu@ictp.it}

\author{Utpal Sarkar$^{2}_{}$}
\email{utpal@prl.res.in}

\affiliation{ $^{1}_{}$The Abdus Salam International Centre for
Theoretical Physics, Strada Costiera 11, 34014 Trieste, Italy
\\
$^{2}_{}$Physical Research Laboratory, Ahmedabad 380009, India}

\begin{abstract}

We propose an extension of the standard model, in which neutrinos
are Dirac particles and their tiny masses originate from a one-loop
radiative diagram. The new fields required by the neutrino
mass-generation also accommodate the explanation for the
matter-antimatter asymmetry and dark matter in the universe.

\end{abstract}


\maketitle

\emph{Introduction:} Various neutrino oscillation experiments
\cite{pdg2006} have confirmed that neutrinos have tiny but nonzero
masses. This phenomenon is naturally explained by the seesaw
mechanism \cite{minkowski1977}. In the original seesaw scenario
neutrinos are assumed to be Majorana particles, whose existence has
not been experimentally verified so far. As an alternative, Dirac
seesaw was proposed \cite{rw1983,gh2006} where the neutrinos can
naturally acquire small Dirac masses. In the seesaw models, the
observed matter-antimatter asymmetry in the universe can also be
generated through the leptogenesis
\cite{fy1986,luty1992,fps1995,pilaftsis1997,ms1998,dlrw1999}.
Another big challenge to the standard model (SM) is the nature of
the dark matter, which contributes about $25\%$ \cite{pdg2006} to
the energy density of the unverse. This also indicates the necessity
of supplementing to the existing theory with newer particles having
GeV order mass and very weak interactions.

In this work, we present a new scenario for a naturally tiny Dirac
neutrino mass, which accounts for the dark matter and accommodate
the leptogeneis. In our model, the small neutrino mass is elegantly
induced through a radiative diagram. The new fields responsible for
the neutrino mass generation also accommodate the CP-violation and
out-of-equilibrium decays to realize the leptogenesis as well as the
candidates for the cold dark matter.

\vskip .2in

\emph{Our model:} We extend the $SU(3)_c^{} \times SU(2)_L^{} \times
U(1)_Y^{}$ SM by introducing two complex scalars: $\chi
(\textbf{1},\textbf{1},0)$, $\eta (\textbf{1},\textbf{2},-1)$, one
real scalar: $\sigma (\textbf{1},\textbf{1},0)$, three fermions:
$S_{L,R}^{} (\textbf{1},\textbf{1},0)$ and three right-handed
neutrinos: $\nu_R^{} (\textbf{1},\textbf{1},0)$. Among these new
fields, we appoint the lepton number of the SM leptons for
$\nu_{R}^{}$ and $S_{L,R}^{c}$, furthermore, we impose a
$U(1)_{D}^{}$ gauge symmetry, under which $\chi$, $S_R^{}$ and
$\nu_{R}^{c}$ carry the quantum number $1$, and a $Z_2^{}$ discrete
symmetry, under which $\eta$, $\sigma$ and $S_{L,R}^{}$ are odd. Our
model exactly conserves the lepton number as well as the
$U(1)_{D}^{}$ and $Z_{2}^{}$, so the allowed interactions involving
$S_{L,R}^{}$ and $\nu_{R}^{}$ are given by
\begin{eqnarray}
\label{lagrangian1} \mathcal{L}\,\supset
-\,y\,\overline{\psi_L^{}}\,\eta\, S_L^{c}\,-\,h\,\sigma
\,\overline{\nu_R^{}}\,S_R^{c}\,- \,f\,\chi\,\overline{S_R^{}}\,
S_L^{}\,+\,\textrm{h.c.}\,,
\end{eqnarray}
where $\psi_{L}^{}(\textbf{1},\textbf{2},-1)$ denotes the SM
left-handed leptons. We also write down the general scalar
potential,
\begin{eqnarray}
\label{potential1} V&=&m_{\chi}^{2}\chi^\dagger_{}\chi
+m_{\phi}^{2}\phi^\dagger_{}\phi+m_{\eta}^{2}\eta^\dagger_{}\eta+\frac{1}{2}m_{\sigma}^{2}\sigma^2_{}\nonumber\\
&&+\lambda_{\chi}^{}\left(\chi^\dagger_{}\chi\right)^{2}_{}+\lambda_{\phi}^{}\left(\phi^\dagger_{}\phi\right)^{2}_{}
+\lambda_{\eta}^{}\left(\eta^\dagger_{}\eta\right)^{2}_{}+\frac{1}{4}\lambda_{\sigma}^{}\sigma^4_{}\nonumber\\
&&+\lambda_{\chi\phi}^{}\chi^\dagger_{}\chi\phi^\dagger_{}\phi
+\lambda_{\chi\eta}^{}\chi^\dagger_{}\chi\eta^\dagger_{}\eta
+\frac{1}{2}\lambda_{\chi\sigma}^{}\chi^\dagger_{}\chi\sigma^2_{}\nonumber\\
&& +\lambda_{\phi\eta}^{}\phi^\dagger_{}\phi\eta^\dagger_{}\eta
+\lambda'^{}_{\phi\eta}\phi^\dagger_{}\eta\eta^\dagger_{}\phi
+\frac{1}{2}\lambda_{\phi\sigma}^{}\phi^\dagger_{}\phi\sigma^2_{}
\nonumber\\
&&+\lambda_{\eta\sigma}^{}\eta^\dagger_{}\eta\sigma^2_{} +\left[
\kappa\left(\phi^\dagger_{}\eta\right)^{2}_{}+\frac{1}{\sqrt{2}}\mu\sigma\eta^\dagger_{}\phi+\textrm{h.c.}\right]\,.
\end{eqnarray}
Here $\phi (\textbf{1},\textbf{2},-1)$ is the SM Higgs. In the
following, we will choose the Yukawa coupling $f$ in
(\ref{lagrangian1}) to be real and diagonal while the quartic
coupling $\kappa$ and the cubic coupling $\mu$ in (\ref{potential1})
to be real for convenience but without loss of generality.

The $Z_2^{}$ symmetry is unbroken at all energies and hence $\eta$
and $\sigma$ are protected from any nonzero vacuum expectation
values ($vev$s). The gauge symmetry $U(1)_{D}^{}$ is expected to
break by $\langle\chi\rangle=\mathcal{O}(10^{9}\,\textrm{GeV})$. In
consequence, the corresponding gauge boson $Z'$ obtain its mass
$M_{Z'}^{}=\sqrt{2}\,g'\,\langle\chi\rangle$ with $g'$ being the
$U(1)$ gauge coupling, meanwhile, $S_{L,R}^{}$ realize their Dirac
masses $M_S^{}=f\langle\chi\rangle$ of the order of
$10^{7-8}\,\textrm{GeV}$. Subsequently, the electroweak symmetry is
broken by $\langle\phi\rangle\simeq174\,\textrm{GeV}$.

\vskip .2in

\begin{figure}
\vspace{6.1cm} \epsfig{file=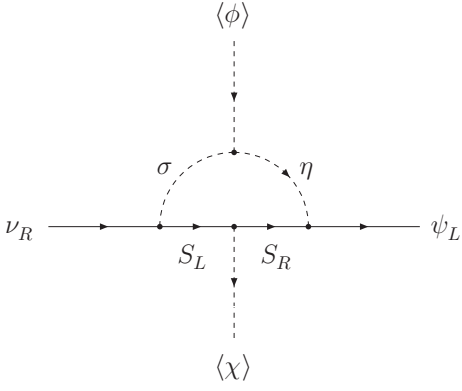, bbllx=5.3cm,
bblly=6.0cm, bburx=15.3cm, bbury=16cm, width=7cm, height=7cm,
angle=0, clip=0} \vspace{-7.5cm} \caption{\label{massgeneration} The
one-loop diagram for generating the radiative neutrino masses.}
\end{figure}

\emph{Neutrino mass:} As shown in Fig. \ref{massgeneration}, the
neutrinos can get a Dirac mass through the one-loop diagram after
the $U(1)_D^{}$ and electroweak symmetry breaking. For
demonstration, we define $\eta^0_{}\equiv
\frac{1}{\sqrt{2}}\left(\eta_{R}^{0}+i\eta_{I}^{0}\right)$, where
$\eta^0_{}$ is the neutral component of $\eta$, and then have
\begin{eqnarray}
\mathcal{L}&\supset&-\,\frac{1}{2}\left(
 \eta^0_{I}~~\eta^0_R~~\sigma\right)\left(
\begin{array}{ccc}
 M_{\eta^0_I}^{2}& 0&0\\[2mm]
 0 & M_{\eta^0_R}^{2}&\Delta_{}^{2}\\[2mm]
 0&\Delta_{}^{2}& M_{\sigma}^{2}\end{array}
\right)\left(\begin{array}{c}
 \eta_{I}^{0} \\[2mm]
 \eta^0_{R}\\[2mm]
 \sigma
\end{array}\right)\,,
\end{eqnarray}
where
\begin{eqnarray}
M_{\eta_I^0}^{2}
&\equiv&m_{\eta}^2\,+\,\lambda_{\chi\eta}^{}\,\langle\chi\rangle^{2}_{}\,
+\,\left(\lambda_{\phi\eta}^{}\,+\lambda'^{}_{\phi\eta}\,-\,\kappa\right)\langle\phi\rangle^{2}_{}\,,\\[2mm]
M_{\eta_R^0}^{2}
&\equiv&m_{\eta}^2\,+\,\lambda_{\chi\eta}^{}\,\langle\chi\rangle^{2}_{}\,
+\,\left(\lambda_{\phi\eta}^{}\,+\lambda'^{}_{\phi\eta}\,+\,\kappa\right)\langle\phi\rangle^{2}_{}\,,\\[2mm]
M_{\sigma}^{2}
&\equiv&m_{\sigma}^2\,+\,\lambda_{\chi\sigma}^{}\,\langle\chi\rangle^{2}_{}\,
+\,\lambda_{\phi\sigma}^{}\,\langle\phi\rangle^{2}_{}\,,\\[2mm]
\Delta_{}^{2} &\equiv& \mu\,\langle\phi\rangle\,.
\end{eqnarray}
In the following, we will take $\kappa>0$ and then
$M_{\eta_I^0}^{}<M_{\eta_R^0}^{}$ for illustration. There is a
simple transformation of $\sigma$ and $\eta^0_{R}$ to the mass
eigenstates $\xi_{1}^{}$ and $\xi_{2}^{}$,
\begin{eqnarray}
\left(\begin{array}{c}
 \eta^0_{R} \\[2mm]
 \sigma \end{array}\right)&=& \left(
\begin{array}{cc}
 \cos\vartheta & \sin\vartheta\\[2mm]
 -\sin\vartheta & \cos\vartheta\end{array}
\right)\left(\begin{array}{c}
 \xi_{1}^{} \\[2mm]
 \xi_{2}^{} \end{array}\right)\,,
\end{eqnarray}
where the mixing angle $\vartheta$ is
\begin{eqnarray}
\label{mixing1}
\tan2\vartheta&=&\frac{2\Delta_{}^{2}}{M_{\sigma}^{2}-M_{\eta^0_R}^{2}}\,.
\end{eqnarray}
The masses of the eigenstates are
\begin{eqnarray}
\label{mass1}
M_{\xi_1}^{2}&=&\frac{1}{2}\left[M_{\eta^0_R}^{2}+M_{\sigma}^{2}-\sqrt{\left(M_{\eta^0_R}^{2}-M_{\sigma}^{2}\right)^{2}_{}
+4\Delta_{}^{4}}\right]\,,\\[2mm]
M_{\xi_2}^{2}&=&\frac{1}{2}\left[M_{\eta^0_R}^{2}+M_{\sigma}^{2}+\sqrt{\left(M_{\eta^0_R}^{2}-M_{\sigma}^{2}\right)^{2}_{}
+4\Delta_{}^{4}}\right]\,.
\end{eqnarray}

We then give the formula of the radiative neutrino masses,
\begin{eqnarray}
\label{mass1}
\left(m_{\nu}^{}\right)_{ij}^{}&=&\frac{\sin2\vartheta}{32\pi^{2}_{}}\,\sum_k^{}\,y_{ik}^{}\,
M_{S_k}^{}\,\left[\frac{M_{\xi_2}^{2}}{M_{\xi_2}^{2}\,-\,M_{S_k}^{2}}
\,\ln
\left(\frac{M_{\xi_2}^{2}}{M_{S_k}^{2}}\right)\right.\,\nonumber\\[2mm]
&&\left.-\,\frac{M_{\xi_1}^{2}}{M_{\xi_1}^{2}\,-\,M_{S_k}^{2}} \,\ln
\left(\frac{M_{\xi_1}^{2}}{M_{S_k}^{2}}\right)\right]
\,h^{\dagger}_{kj}\,.
\end{eqnarray}
For $M_{S_k}^{}\gg M_{\xi_{1,2}}^{}$, we can simplify the above mass
matrix as
\begin{eqnarray}
\label{mass2}
\left(m_{\nu}^{}\right)_{ij}^{}\simeq\frac{\sin2\vartheta}{32\pi^{2}_{}}\,M_{\xi_1}^{2}\,\sum_k^{}\,
F\left(\frac{M_{\xi_2}^{2}}{M_{\xi_1}^{2}},\frac{M_{S_k}^{2}}{M_{\xi_1}^{2}}\right)
\frac{y_{ik}^{}\,h^{\dagger}_{kj}}{M_{S_k}^{}}\,,
\end{eqnarray}
with the definition
\begin{eqnarray}
\label{mass3} F(x,y) &\equiv&x\,\ln \left(\frac{y}{x}\right)\,-\,\ln
y\,.
\end{eqnarray}
The function
$\displaystyle{F\left(\frac{M_{\xi_2}^{2}}{M_{\xi_1}^{2}},\frac{M_{S_k}^{2}}{M_{\xi_1}^{2}}\right)}$
could be simply looked on as a constant $c$ if the three
$M_{S_k}^{}$ are chosen within a few order of magnitude and then the
neutrino mass can be written in the simple form,
\begin{eqnarray}
\label{mass4}
m_{\nu}^{}&\simeq&c\,\frac{\sin2\vartheta}{32\pi^{2}_{}}\,M_{\xi_1}^{2}\,y\,\frac{1}{M_{S}^{}}\,h^{\dagger}_{}\,.
\end{eqnarray}

For the purpose of numerical estimation, we input
$(M_{\eta^0_R}^{},\,M_{\sigma}^{},\,|\Delta|)=(300\,\textrm{Gev},\,80\,\textrm{Gev},\,30\,\textrm{Gev})\,
\textrm{or}\,(80\,\textrm{Gev},\,300\,\textrm{Gev},\,30\,\textrm{Gev})$
and then obtain
$(\tan2\vartheta,\,M_{\xi_1}^{},\,M_{\xi_2}^{})=(\mp\,0.02,\,80\,\textrm{GeV},\,300\,\textrm{GeV})$.
Subsequently, we take $M_{S_k}^{}\simeq 10^7_{}\,\textrm{GeV}$ and
then derive $c\simeq 270$. For $y\sim h\sim
\mathcal{O}(10^{-3}_{})$, the neutrino mass comes out to be of the
order of $\mathcal{O}(0.01-0.1\,\textrm{eV})$, which is consistent
with the neutrino oscillation data and cosmological observations.

\vskip .2in

\emph{Dark matter:} It is natural to consider the lighter one
between $\xi_{1}^{}$ and $\eta^0_I$ as the candidate for the dark
matter since it has not any decay modes. We first consider that
$\xi_1^{}$ and $\xi_2^{}$ are dominated by $\sigma$ and $\eta^0_R$,
respectively and $\sigma$ is lighter than $\eta^0_I$. In this case,
$\sigma$ is definitely the darkon field \cite{sz1985} that can
realize the right amount of the relic density of the cold dark
matter when its mass $M_{\sigma}^{}$ is less than
$100\,\textrm{GeV}$ and its quartic coupling to the SM Higgs, i.e.
$\lambda_{\phi\sigma}^{}$ in the scalar potential (\ref{potential1})
is of the order of $\mathcal{O}(0.1)$. Now we check the other
possibility that $\eta^0_I$ is the dark matter. Note that the direct
detection of halo dark matter places a limit on the mass degeneracy
between $\eta^0_R$ and $\eta^0_I$, because the difference must be
sufficient to kinematically suppress the scattering of
$\eta^0_{R,I}$ on nuclei via the tree-level exchange of the $Z$
boson. It has been studied \cite{bhr2006} that if $\eta^0_I$ is
expected to be the dark matter, the mass spectrum of
mass-eigenstates $\eta^0_{R,I}$ should be:
\begin{eqnarray}
M_{\eta^0_R}^{}-M_{\eta^0_I}^{} \simeq (8-9)\,\textrm{GeV}\,
\end{eqnarray}
for
\begin{eqnarray}
M_{\eta^0_I}^{}=(60-73)\,\textrm{GeV}\,,
\end{eqnarray}
or
\begin{eqnarray}
M_{\eta^0_R}^{}-M_{\eta^0_I}^{}\simeq (9- 12)\,\textrm{GeV}
\end{eqnarray}
for
\begin{eqnarray}
M_{\eta_I^0}^{}=(73- 75)\,\textrm{GeV}\,.
\end{eqnarray}
In our model, we have the flexibility to choose the quartic coupling
$\kappa$ and other parameters in the scalar potential
(\ref{potential1}) and then obtain the desired mass spectrum of
$\eta^0_{R,I}$.

\vskip .2in

\emph{Leptogenesis:} Obviously, no lepton asymmetry can be generated
in our model because the lepton number is exactly conserved.
However, since the sphaleron \cite{krs1985} only have a direct
action on the left-handed quarks and leptons, a nonzero lepton
asymmetry stored in the left-handed leptons, which is equal but
opposite to that stored in the other fields, can be partially
converted to the baryon asymmetry as long as the interactions
between the left-handed leptons and the other fields with lepton
number are too weak to realize an equilibrium before the electroweak
phase transition. For all the SM species, the Yukawa interactions
are sufficiently strong to rapidly cancel the left- and right-handed
lepton asymmetry. But the effective Yukawa interactions of the
ultralight Dirac neutrinos are exceedingly weak and thus will not
reach equilibrium until the temperatures fall well below the weak
scale. This new type of leptogenesis mechanism is called
neutrinogenesis \cite{dlrw1999}.

\begin{figure}
\vspace{7.0cm} \epsfig{file=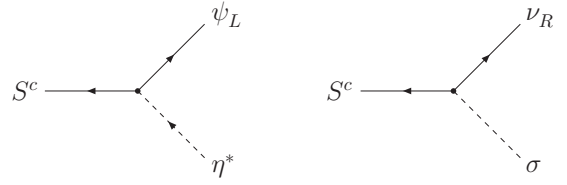, bbllx=3cm, bblly=6.0cm,
bburx=13cm, bbury=16cm, width=7cm, height=7cm, angle=0, clip=0}
\vspace{-10cm} \caption{\label{decaytree} The heavy Dirac fermions
decay to the left-handed leptons and the right-handed neutrinos.}
\end{figure}

\begin{figure}
\vspace{7.5cm} \epsfig{file=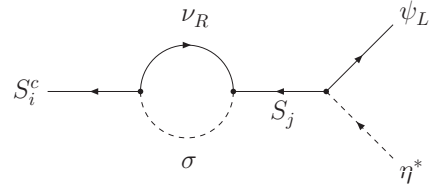, bbllx=0.5cm, bblly=6.0cm,
bburx=10.5cm, bbury=16cm, width=7cm, height=7cm, angle=0, clip=0}
\vspace{-8cm} \caption{\label{decayloop} The heavy Dirac fermions
decay to the left-handed leptons at one-loop order. }
\end{figure}

In our model, the heavy Dirac fermions $S=S_{L}^{}+S_R^{}$ have two
decay modes as shown in Fig. \ref{decaytree}. We calculate the decay
width at tree level,
\begin{eqnarray}
\label{decaywidth11} &&\Gamma\left(S_i^{c}\rightarrow
\psi_L^{}+\eta^\ast_{}\right)=\Gamma\left(S_i^{}\rightarrow
\psi_L^{c}+\eta\right)\\
&=&\frac{1}{16\pi}\left(y^{\dagger}_{}y\right)_{ii}^{}\,M_{S_i}^{}\,,\\
[4mm] &&\Gamma\left(S_i^{c}\rightarrow
\nu_R^{}+\sigma\right)=\Gamma\left(S_i^{}\rightarrow
\nu_R^{c}+\sigma\right)\nonumber\\
&=&\frac{1}{32\pi}\left(h^{\dagger}_{}h\right)_{ii}^{}\,M_{S_i}^{}\,.
\end{eqnarray}
At one-loop order as shown in Fig. \ref{decayloop}, we compute the
CP asymmetry
\begin{eqnarray}
\label{cpasymmetry11} \varepsilon_{S_i}^{}&\equiv & \frac{\Gamma
(S_i^{c}\rightarrow \psi_L^{}+\eta^\ast_{})- \Gamma
(S_i^{}\rightarrow
\psi_L^{c}+\eta)}{\Gamma_{S_i}^{}}\nonumber\\[1mm]
&=&\frac{1}{8\pi}\frac{1}{\left(y^{\dagger}_{}y\right)_{ii}^{}\,+\,\frac{1}{2}\,\left(h^{\dagger}_{}h\right)_{ii}^{}}\nonumber\\[2mm]
&&\times\,\sum_{j\neq
i}^{}\textrm{Im}\left[\left(y^{\dagger}_{}y\right)_{ij}^{}\left(h^{\dagger}_{}h\right)_{ji}^{}\right]
\,\frac{M_{S_i}^{}M_{S_j}^{}}{M_{S_i}^{2}-M_{S_j}^{2}}\,.
\end{eqnarray}
Here the total decay width $\Gamma_{S_i}^{}$ is given by
\begin{eqnarray}
\label{totaldecaywidth11}
\Gamma_{S_i}^{}&=&\frac{1}{16\pi}\left[\left(y^{\dagger}_{}y\right)_{ii}^{}\,
+\,\frac{1}{2}\,\left(h^{\dagger}_{}h\right)_{ii}^{}\right]\,M_{S_i}^{}\,.
\end{eqnarray}

For illustration, we consider the limiting case with $M_{S_1}^{}\ll
M_{S_{2,3}}^{}$, where the final lepton asymmetry should mainly come
from the contributions of the decays of $S_1^{}$. We can simplify
the CP asymmetry (\ref{cpasymmetry11}) as
\begin{eqnarray}
\label{cpasymmetry12} \varepsilon_{S_1}^{}
&\simeq&-\,\frac{1}{8\pi}\frac{1}{\left(y^{\dagger}_{}y\right)_{11}^{}\,+\,\frac{1}{2}\,\left(h^{\dagger}_{}h\right)_{11}^{}}
\nonumber\\[2mm]
&&\times\,\sum_{j\neq
1}^{}\,\frac{M_{S_1}^{}}{M_{S_j}^{}}\textrm{Im}\left[\left(y^{\dagger}_{}y\right)_{1j}^{}\left(h^{\dagger}_{}h\right)_{j1}^{}\right]\,.
\end{eqnarray}
Furthermore, we take a simple assumption,
\begin{eqnarray}
\label{assumption11}
 h&=&y^\ast_{}\,,
\end{eqnarray}
and then approach
\begin{eqnarray}
\label{cpasymmetry13} \varepsilon_{S_1}^{}
&\simeq&-\,\frac{1}{8\pi}\frac{1}{\left(y^{\dagger}_{}y\right)_{11}^{}}\sum_{j\neq
1}^{}\,\frac{M_{S_1}^{}}{M_{S_j}^{}}\,\left\{\textrm{Im}\left[\left(y^{\dagger}_{}y\right)_{1j}^{2}\right]\right\}\,.
\end{eqnarray}
Similar to the DI bound \cite{di2002}, we can also deduce an upper
bound on the above CP asymmetry by inserting the assumption
(\ref{assumption11}) to the mass formula (\ref{mass2}),
\begin{eqnarray}
\left|\varepsilon_{S_1}^{}\right|&\leq&\frac{4\pi}{c \sin
2\vartheta}\frac{M_{S_1}^{}m_3^{}}{M_{\xi_1}^{2}}|\sin\delta|\,,
\end{eqnarray}
with $m_3^{}$ and $\delta$ being the biggest eigenvalue of the
neutrino mass matrix and the CP phase, respectively. Here we have
assumed the neutrinos to be hierarchical \cite{sv2006}. Inputting
$M_{S_1}^{}=10^7_{}\,\textrm{GeV}$, $m_3^{}=0.05\,\textrm{eV}$,
$M_{\xi_1}^{}=80\,\textrm{GeV}$, $c=270$,
$\sin2\vartheta\simeq\tan2\vartheta=0.02$ and $\sin\delta=-1$, we
derive $\varepsilon_{S_1}^{}\simeq -1.8\times 10^{-7}_{}$ and then
obtain the final baryon asymmetry,
\begin{eqnarray}
\label{baryonasymmetry11}
\frac{n_B^{}}{s}&=&\frac{28}{79}\,\frac{n_{B-L_{SM}^{}}^{}}{s}=-\frac{28}{79}\,\frac{n_{L_{SM}^{}}^{}}{s}\nonumber\\
&\simeq&
-\frac{28}{79}\,\varepsilon_{S_1}^{}\frac{n_{S_1}^{eq}}{s}\left|_{T=M_{S_1}^{}}^{}\right.\simeq
-\frac{1}{15}\frac{\varepsilon_{S_1}^{}}{g_\ast^{}}\simeq 10^{-10}
\end{eqnarray}
as desired to explain the matter-antimatter asymmetry of the
universe. Here we have adopted the relativistic degrees of freedom
$g_{\ast}^{}= \mathcal{O}(100)$ \cite{kt1990}.

Note when generating the desired baryon asymmetry
(\ref{baryonasymmetry11}), the decays of $S_1^{}$ should satisfy the
condition of departure from equilibrium, which is described by
\begin{eqnarray}
\label{condition} \Gamma_{S_1}^{}\lesssim
H(T)\left|_{T=M_{S_1}^{}}^{}\right.\,,
\end{eqnarray}
where
\begin{eqnarray}
\label{hubble}
H(T)&=&\left(\frac{8\pi^{3}_{}g_{\ast}^{}}{90}\right)^{\frac{1}{2}}_{}\frac{T^{2}_{}}{M_{\textrm{Pl}}^{}}
\end{eqnarray}
is the Hubble constant with the Planck mass
$M_{\textrm{Pl}}^{}\simeq 10^{19}_{}\,\textrm{GeV}$. With Eqs.
(\ref{totaldecaywidth11}), (\ref{assumption11}), (\ref{condition})
and (\ref{hubble}), it is straightforward to perform the condition
\begin{eqnarray}
\left(y^\dagger
y\right)_{11}^{}&\lesssim&\left(\frac{2^{10}_{}\cdot\pi^5_{}\cdot
g_\ast^{}}{5\cdot
3^4}\right)^{\frac{1}{2}}_{}\,\frac{M_{S_1}^{}}{M_{\textrm{Pl}}^{}}\sim
10^{-10}_{}
\end{eqnarray}
for $M_{S_1}^{}=10^7_{}\,\textrm{GeV}$.

\vskip .2in

\emph{Summary:} In this paper, we extended the SM with the
requirement of the symmetry that forbids the usual Dirac and
Majorana masses of the neutrinos. Through a radiative diagram, the
neutrinos obtain tiny Dirac masses suppressed by the heavy new
fermions. These fermions also generate a lepton asymmetry stored in
the left-handed leptons via their CP-violation and
out-of-equilibrium decays. The sphaleron action then partially
transfers this lepton asymmetry to a baryon asymmetry so that the
observed matter-antimatter asymmetry of the universe can be
naturally explained. Moreover, the scalars contributing to the
neutrino mass-generation provide consistent candidates for the cold
dark matter.

\end{document}